\documentclass[10pt,aps,twocolumn,prd,superscriptaddress,noshowpacs,nofootinbib,noshowkeys,floatfix]{revtex4}
\usepackage[dvips]{graphics,graphicx}
\usepackage{amsmath, amssymb}
\usepackage{multirow}
\usepackage{longtable}
\usepackage{color}
\usepackage[normalem]{ulem}  

\renewcommand\sout{\bgroup \color{red} \ULdepth=-.5ex \ULset}

\begin{document}
\title{ Confronting fluctuations of conserved charges in central nuclear
  collisions at the LHC with predictions from Lattice QCD }
\date{\today}
\author{P. Braun-Munzinger}
\affiliation{Extreme Matter Institute EMMI, GSI,
Planckstr. 1, D-64291 Darmstadt, Germany}
\affiliation{Technische Universit\"at Darmstadt, Darmstadt, Germany}
\affiliation{Frankfurt Institute for Advanced Studies, J.W. Goethe Universit\"at, Frankfurt, Germany}
\author{A. Kalweit}
\affiliation{European Organization for Nuclear Research (CERN), Geneva, Switzerland}
\author{K. Redlich}
\affiliation{Institute of Theoretical Physics, University of Wroclaw,
PL-50204 Wroc\l aw, Poland}
\affiliation{Extreme Matter Institute EMMI, GSI,
Planckstr. 1, D-64291 Darmstadt, Germany}
\author{J. Stachel}
\affiliation{Physikalisches Institut, Universit\"at Heidelberg, Heidelberg, Germany}

\begin{abstract}

We construct net baryon number and strangeness susceptibilities as
well as correlations between electric charge, strangeness and baryon
number from experimental data at midrapidity of the ALICE Collaboration at CERN. The
data were taken in central Pb-Pb collisions at $\sqrt{s_{NN}}$=2.76
TeV and cover one unit of rapidity.   The
resulting fluctuations and correlations are consistent with Lattice
QCD results at the chiral crossover pseudocritical temperature
$T_c\simeq 155$ MeV. This agreement lends strong support to the
assumption that the fireball created in these collisions is of thermal
origin and exhibits characteristic properties expected in QCD at the
transition from the quark gluon plasma to the hadronic phase. The
volume of the fireball for one unit of rapidity at $T_c$ is found to
exceed 3000 fm$^3$. A detailed discussion on uncertainties in the
temperature and volume of the fireball is presented. The results are
linked to pion interferometry measurements and predictions from
percolation theory.
\end{abstract}
\maketitle

\section{Introduction}

Uncovering evidence for (partial) restoration of chiral symmetry in
the medium created in nucleus-nucleus collisions at very high energy
is one of the most important but also challenging problems
\cite{friman11:_cbm_physic_book,fukushima11:_phase_diagr_of_dense_qcd,Fukushima:2013rx}.
Recently, experimental studies along this line have been carried out
by measuring fluctuations of conserved charges
\cite{aggarwal10:_higher_momen_of_net_proton,STAR_pn_2013,adamczyk14:_beam_energ_depen_of_momen}
as part of the RHIC Beam Energy Scan (BES) program.

Fluctuations of conserved charges are particularly interesting probes
of critical phenomena and the phase diagram in QCD
\cite{hatta03:_proton_number_fluct_as_signal,stephanov98:_signat_of_tricr_point_in_qcd,ejiri06:_hadron_fluct_at_qcd_phase_trans,stephanov09:_non_gauss_fluct_near_qcd_critic_point,stephanov11:_sign_of_kurtos_near_qcd_critic_point},
     {as well as freezeout conditions in heavy ion collisions}
     \cite{karsch11:_probin_freez_out_condit_in,Bazavov:2012vg,Ratti,cratti}.
     The intent of the present work is to provide a link between
      fluctuations derived from measurements of particle yields in Pb--Pb collisions at the LHC and predictions
     from Lattice QCD.

Early on, the QCD phase transition was conjectured to be of second
order, belonging to the $O(4)$ universality class
\cite{pisarski84:_remar_chiral_trans_in_chrom}, for massless light
quarks.  Current Lattice QCD (LQCD) simulations at physical quark
masses show that, at vanishing or small baryon density, the transition
from a hadron gas to a quark gluon plasma is most likely a crossover
\cite{aoki06}. The results further indicate that the chiral crossover
appears in the critical region of the second order transition
belonging to the {O(2)/O(4)} universality class
\cite{ejiri09:_magnet_equat_of_state_in_flavor_qcd,kaczmarek11:_phase_qcd,allton05:_therm_of_two_flavor_qcd}. Consequently,
observables such as fluctuations of net baryon number and electric
charge, which are sensitive to criticality related with a spontaneous
breaking of chiral symmetry, should exhibit characteristic properties
governed by the universal part of the free energy
\cite{karsch11:_probin_freez_out_condit_in,ejiri06:_hadron_fluct_at_qcd_phase_trans,friman11:_fluct_as_probe_of_qcd}.

The magnetic equation of state and cumulants of net charges at
physical quark masses have been studied in LQCD calculations
\cite{allton05:_therm_of_two_flavor_qcd,bazavov12:_fluct_and_correl_of_net,borsanyi12:_fluct_of_conser_charg_at,cheng09:_baryon_number_stran_and_elect,Bazavov:2014xya},
as well as in effective chiral models
\cite{fukushima04:_chiral_polyak,sasaki07:_quark_number_fluct_in_chiral,sasaki07:_suscep_polyakov,stokic09:_kurtos_and_compr_near_chiral_trans,skokov10:_meson_fluct_and_therm_of,skokov10:_vacuum_fluct_and_therm_of_chiral_model,skokov11:_quark_number_fluct_in_polyak,friman11:_fluct_as_probe_of_qcd,asakawa09:_third_momen_of_conser_charg,herbst11:_phase_struc_of_polyak_quark,schaefer12:_qcd_critic_region_and_higher,wagner10:_effic_comput_of_high_order}.
Their properties have been shown to be consistent with general
expectations for $O(4)$ scaling.
These results have opened a new approach to get experimental
information on the QCD phase boundary,  by measuring higher moments of
distributions of event-by-event fluctuations of conserved charges
\cite{karsch11:_probin_freez_out_condit_in,braun-munzinger11:_net_proton_probab_distr_in,
kaczmarek11:_phase_qcd,bazavov12:_fluct_and_correl_of_net,friman11:_fluct_as_probe_of_qcd,Bazavov:2012vg},
and their probability distributions
\cite{BraunMunzinger:2011ta,morita12:_baryon_number_probab_distr_near,morita13:_net}.

The direct measurement of higher moments of event-by-event
fluctuations is complicated by several issues. First, quantities like
fluctuations of the net baryon number can only be reliably obtained if
effective methods are applied to
correct the data for
the efficiency of the detector. Furthermore, for conserved quantities
like baryon number, appropriate corrections need to be applied, due to
the finite detector acceptance. The situation has been analyzed by
\cite{bzdak12:_accep_correc_to_net_baryon,bzdak13:_baryon_number_conser_and_cumul} and attempts at corrections have been applied for
acceptance
\cite{STAR_pn_2013,adamczyk14:_beam_energ_depen_of_momen}
and for fluctuations induced by the difficult to measure neutral
baryons
\cite{kitazawa12:_reveal_baryon_number_fluct_from,kitazawa04:_relat_between_baryon_number_fluct}. Second,
such measurements are sensitive to critical effects near the QCD phase
boundary only for higher moments of the distributions
\cite{friman11:_fluct_as_probe_of_qcd}, necessitating huge statistics
as well as a very precise understanding of possible backgrounds in the
measurements.  Here we present a different approach, where the
second order cumulants and correlations of conserved charges
are directly obtained from the measured
inclusive distributions, albeit under a special assumption on the
shape of the probability distributions.

In the special case that the probability distribution of the number of
particles $N_q$ and antiparticles $N_{-q}$ is uncorrelated and
Poisson, the probability distribution of the variable $N=N_q-N_{- q}$
is a Skellam function, which is entirely determined by the mean number
of particles $\langle N_q\rangle$ and antiparticles $\langle
N_{-q}\rangle$ \cite{braun-munzinger11:_net_proton_probab_distr_in}. As is explained in the
following section one can then determine the 2nd order
susceptibilities directly from inclusive measurements.

The assumption of a Skellam distribution for the distribution of net
baryon number seems to be well fulfilled at RHIC energies {up to the
  second order cumulants}
\cite{STAR_pn_2013,adamczyk14:_beam_energ_depen_of_momen,braun-munzinger11:_net_proton_probab_distr_in},
see also the discussion below. Nevertheless, assuming independent
production of baryons and anti-baryons is a strong
assumption. However, one should note that generalized susceptibilities
$c_B^n$ do not contain, at least for $\mu_B = 0$, any singular terms
corresponding to chiral critical behavior if $n < 6$
\cite{ejiri06:_hadron_fluct_at_qcd_phase_trans,friman11:_fluct_as_probe_of_qcd}. Further,
there is strong evidence that the fireball is very close to thermal
equilibrium {as demonstrated by analysis within the framework of
  the Hadron Resonance Gas (HRG) partition function}
\cite{BraunMunzinger:1999qy,BraunMunzinger:2001ip,Statmodelreview_QGP3,Andronic:2011yq,Andronic:2012dm,Stachel:2013zma,Becattini:2014hla},
     {which also quantifies the LQCD equation of state in the confined
       phase}
     \cite{bazavov12:_fluct_and_correl_of_net,Karsch:2003zq,Karsch:2013naa}.

     The current approach leads then to a direct connection between
     experimental data integrated over all transverse momenta and
     second order susceptibilities and, consequently, to direct
     contact between predictions from LQCD and experimental data
     without the need to consider, on the experimental side, effects
     of acceptance and, on the theoretical side, how to extract baryons
     from LQCD calculations.

\section{Fluctuations and correlations of net charges} \label{sec:fluc_corr}

We consider a thermal medium of strongly interacting particles of
volume $V$ at temperature $T$, where the baryon number $B$,
strangeness $S$ and electric charge $Q$ are conserved on the
average. The thermodynamics of such a system is characterized by the
pressure, $P(T,V,\vec \mu)$ in the grand canonical ensemble, where
$\vec \mu=(\mu_B,\mu_S,\mu_Q)$ are chemical potentials which
guarantee the conservation of all 'charges' $q =(B,Q,S)$.

In this thermal medium,  fluctuations   of the  net charge  $N$
 \begin{align}\label{fluctuation}
\hat\chi_{N} \equiv {{\chi_{N}}\over T^2} = {{\partial^2\hat P}\over {\partial \hat
\mu^2_N}},
\end{align}
and  correlations  $\chi_{N,M}$ of charges $N$ and $M$
  \begin{align}\label{correlation}
  \hat\chi_{NM}\equiv  {{\chi_{NM}}\over {T^2}} ={{\partial^2\hat P}\over {\partial \hat
\mu_{N}\partial\hat
\mu_{M} }}
\end{align}
are  obtained as derivatives of the reduced thermodynamic pressure $\hat P=P/T^4$,
with respect to the corresponding reduced chemical potential $\hat
\mu_N=\mu_N/T$, where $N,M=(B,S,Q)$.

 \begin{figure}[!t]
\vskip -1.3cm \includegraphics[width=3.60in,angle=-90]{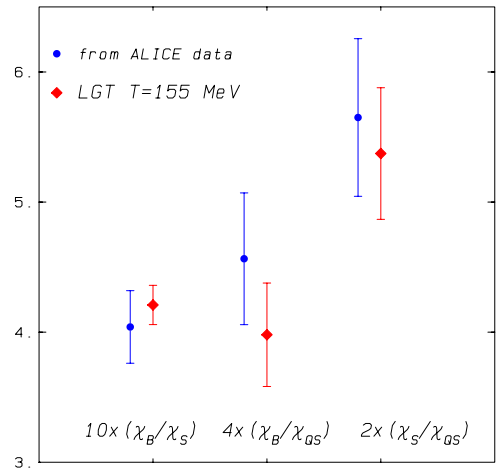}\hskip 0.9cm
 \vskip -0.9cm   \caption{Comparison of different susceptibility
   ratios obtained by using data measured by the ALICE collaboration in Eq.  (\ref{nratio}) with LQCD
   results at $T=155$ MeV from Refs. \cite{bazavov12:_fluct_and_correl_of_net,Bazavov:2014xya}.   }\label{fig1}
 \end{figure}

The susceptibility of a  conserved charge can be also related to its variance,
\begin{align}\label{var}
\hat\chi_{N}  = {1\over {VT^3}}(\langle N^2\rangle -\langle N\rangle^2).
\end{align}
If $P(N)$ is the probability distribution of a conserved charge $N$,
then the $n$-th moment $\langle N^n\rangle $, is calculated as
\begin{align}\label{moment}
\langle N^n\rangle  = \sum_N N^n P(N).
\end{align}

 For the special case of a Skellam distribution, and from Eqs. (\ref{var}) and
 (\ref{moment}), the susceptibility is determined
 by the total mean
 number of particles and antiparticles  \cite{braun-munzinger11:_net_proton_probab_distr_in},
 \begin{align}\label{vars}
 {{\chi_N}\over T^2}=\frac{1}{VT^3}(\langle N_{q}\rangle+\langle N_{-q}\rangle).
\end{align}
The above result is valid if there are only particles of the same
charge, as for baryons, where the charge is $B = 1$. For strangeness and an
electric charge, there are hadrons with charge two and three. In this
case, the Skellam probability distribution can be generalized, and $P(N)$
is expressed by the mean numbers of all particles and antiparticles of
different charges $Q,S$ \cite{BraunMunzinger:2011ta}. The net charge susceptibility is then obtained
from
\begin{align}\label{varsc}
 \hat\chi_N={{\chi_N}\over T^2}=\frac{1}{VT^3}\sum_{n=1}^{|q|} n^2(\langle N_{n}\rangle+\langle N_{-n}\rangle),
\end{align}
where $|q|=(1,2)$ and  $|q|=(1,2,3)$ for electric charge and strangeness, respectively.

For the correlation of different charges, the corresponding expression  reads
  \begin{align}\label{cors}
\hat\chi_{NM}= {{\chi_{NM}}\over T^2}=\frac{1}{VT^3}\sum_{n=-q_N}^{q_N} \sum_{m=-q_M}^{q_M}n  m\langle N_{n,m}\rangle,
\end{align}
where  $\langle N_{n,m}\rangle$, is the mean number of particles and resonances  carrying charges
$N=n$ and $M=m$.

\subsection{Modeling susceptibilities and correlations in Heavy Ion Collisions at the LHC}

The probability distribution of fluctuations of conserved charges can, in general, be measured in
heavy ion collisions using event-by-event analysis. The results for
fluctuations of the net baryon, or rather
net proton number, obtained by the STAR
Collaboration at RHIC
\cite{STAR_pn_2013,adamczyk14:_beam_energ_depen_of_momen},
demonstrate clearly
that, in central Au-Au collisions at $\sqrt{s_{NN}} =200$ GeV, the
fluctuations up to  third order can be {well} described by the Skellam distribution.  Thus, for small $N$, the
distribution $P(N)$ of protons and antiprotons must be independent and very close to Poisson.
No dramatic changes in soft particle production have been observed so
far when going from RHIC to LHC energy. Consequently, the assumption
of independent particle production also at LHC energy seems well
founded and, moreover, can be directly tested experimentally.

We take advantage of the above experimental observations, and construct
the fluctuations and correlations in central Pb--Pb collisions at
the LHC by using results of Eqs. (\ref{vars}), (\ref{varsc}) and
(\ref{cors}).  This way we obtain the susceptibilities $\chi_B$, $\chi_S$ and $\chi_{QS}$ from
particle yields, measured by the  ALICE Collaboration at central rapidity.

The net baryon number fluctuations are obtained as

 \begin{align}\label{bsd}
 {{\chi_B}\over T^2} &={1\over{VT^3}}[ \langle p\rangle+ \langle
   N\rangle+\langle\Lambda+\Sigma^0\rangle+\langle\Sigma^+\rangle+\langle\Sigma^-\rangle
   \\\nonumber & +\langle\Xi^-\rangle+ \langle\Xi^0\rangle
   +\langle\Omega^-\rangle + {\rm antiparticles} ],
\end{align}
where $\langle\,\rangle$ denotes  the corresponding mean particle yield per unit rapidity.

The net strangeness susceptibility is calculated following Eq. (\ref{varsc}), and  approximated as
\begin{align}\label{ssd}
 {{\chi_S}\over T^2} &\simeq {1\over{VT^3}}[ ( \langle
   K^+\rangle+\langle
   K^0\rangle+\langle\Lambda+\Sigma^0\rangle+\langle\Sigma^+\rangle
   \\\nonumber &+\langle\Sigma^-\rangle+4\langle\Xi^-\rangle+
   4\langle\Xi^0\rangle+9\langle\Omega^-\rangle+{\rm antiparticles})
   \\\nonumber &- (\Gamma_{\phi\to K^+}+\Gamma_{\phi\to
     K^-}+\Gamma_{\phi\to K^0}+\Gamma_{\phi\to
     {\bar K^0}})\langle\phi\rangle ].
\end{align}

At LHC energy we assume that $K^0 = \bar K^0 = K^+$ and take the experimentally
measured value. In the kaon yields $\langle K \rangle$ , there are contributions from non strange resonances
decaying into kaons. From Eq.  (\ref{fluctuation}), it is clear, that
such particles should not contribute to strangeness fluctuations. To
correct for the above, we have subtracted kaons coming from $\phi$
decay. The contributions of further non-strange resonances cannot be
accounted for since their yields are not known. However, due to their
larger masses, such contributions are subleading.

The mixed susceptibilities, from Eq (\ref{correlation}), are selecting
contributions of particles carrying the corresponding quantum
numbers. We consider, the strangeness-charge correlations
$\chi_{QS}$. Following Eq. (\ref{cors}), the $\chi_{QS}$ receive
contributions only from strange particles with non vanishing electric
charge. We construct strangeness-charge correlations from particle
yields as
\begin{align}\label{corQS}
 {{\chi_{QS}}\over {T^2}} &\simeq {1\over{VT^3}}[ ( \langle K^+\rangle
+2\langle\Xi^-\rangle
+3\langle\Omega^-\rangle \\\nonumber
&+{\rm antiparticles})- (\Gamma_{\phi\to K^+}+\Gamma_{\phi\to K^-})\langle\phi\rangle\\\nonumber
&- (\Gamma_{K_0^*\to K^+}+\Gamma_{K_0^*\to K^-})\langle K_0^*\rangle
 ],
\end{align}
where we have again subtracted the contribution from decays of $\phi$ and
$K_0^*$, which are contributing to charged kaon yields, but according
to Eq. (\ref{correlation}), should not be included. As in the case of
$\chi_S$, there are also decays of further non-strange, as well as,
neutral strange hadrons which are contributing to $\langle K^\pm \rangle$, but
should not be included.
{ However, due to  lack of data
their contribution can not be subtracted;
nevertheless it is expected to be small.}

\begin{figure}[!t]
 \vskip -1.3cm\includegraphics[width=3.60in,angle=-90]{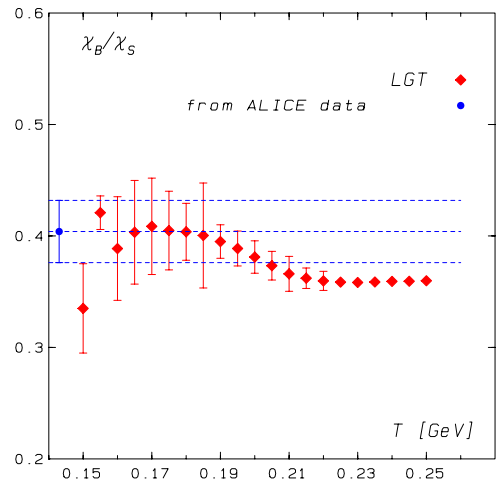}\hskip 0.9cm
 \vskip -0.9cm \caption{The LQCD results on temperature dependent baryon
   number and strangeness susceptibility ratio from Ref.
   \cite{bazavov12:_fluct_and_correl_of_net}.
   The LQCD value at $T=155$ MeV is from Ref. \cite{Bazavov:2014xya}.  Also shown is a band
   for the expected value of this ratio constructed from ALICE data
   in Eq. (\ref{nratio}).  }\label{fig2}
 \end{figure}

In heavy ion collisions at the LHC, due to transparency, particles and
antiparticles are produced symmetrically at midrapidity. Consequently,
the yields of particles and their antiparticle are identical. In
addition, at mid-rapidity, the system is isospin symmetric and charge
neutral, thus leading to equal number of protons and neutrons, and,
more generally, to equal yields for different charge states of the
same particle. Consequently, from Eqs. (\ref{bsd}), (\ref{ssd}) and
(\ref{corQS}), one gets

\begin{align}
&{{\chi_B}\over T^2} = {1\over{VT^3}}[  4\langle p\rangle+ 2\langle(\Lambda+\Sigma^0)\rangle
  + 4\langle\Sigma^{+}\rangle\\\nonumber
 &~~~~~~
  +4\langle\Xi\rangle
+2\langle\Omega\rangle
 ]\\\nonumber
& {{\chi_S}\over T^2} \simeq {1\over{VT^3}}[ 2 \langle K^{+}\rangle+2\langle K^0\rangle
+2\langle(\Lambda+\Sigma^{0})\rangle
  +4\langle\Sigma^{+}\rangle\\
  \nonumber&~~~~~~+16\langle\Xi\rangle+
18\langle\Omega\rangle
- 2(\Gamma_{\phi\to K^+}+\Gamma_{\phi\to K^0})\langle\phi\rangle ]\\\nonumber
& {{\chi_{QS}}\over T^2} \simeq {1\over{VT^3}}[ 2 \langle K^{+}\rangle
+4\langle\Xi^{-}\rangle+
6\langle\Omega^{-}\rangle
- 2\Gamma_{\phi\to K^+}\langle\phi\rangle\\
&~~~~~~- 2\Gamma_{K_0^*\to K^+}\langle K^*\rangle\nonumber
 ].
\protect\label{zall}
\end{align}

Furthermore, from data on inclusive $\Lambda$ and $\Sigma^0$
production in pBe collisions at $\sqrt s = 25$ GeV, we obtain the ratio,
$\Sigma^0/\Lambda=0.278\pm 0.011\pm 0.05$, with a statistical and a
systematic error, respectively \cite{Lambda}. We take
$\Sigma^0/\Lambda=0.278\pm 0.052$, thus
$\langle\Sigma\rangle=(0.2175\pm 0.032)\langle\Lambda+\Sigma^0\rangle$.
The branching ratios, $\Gamma_{\phi\to K^0}=0.342\pm 0.004$,
$\Gamma_{\phi\to K}=0.489\pm 0.005$, and $\Gamma_{K_0^*\to
  K^+}=0.666$ are from  \cite{pdg}.

\begin{figure}[!t]
  \vskip -1.3cm\includegraphics[width=3.60in,angle=-90]{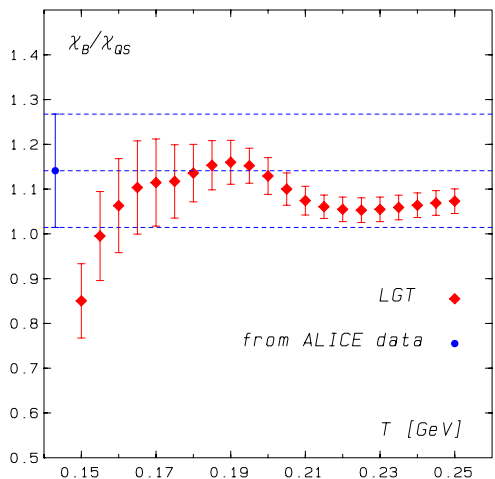}\hskip 0.9cm
 \vskip -1.2cm \caption{Ratio $\chi_B/\chi_{QS}$ from LQCD data from
   Refs. \cite{Bazavov:2014xya,bazavov12:_fluct_and_correl_of_net}, and obtained from
   ALICE data in Eq. (\ref{nratio}).  }\label{fig3}
 \end{figure}

The charge susceptibilities and correlations between conserved charges
can be calculated from the recent ALICE Collaboration data for
particle yields per unit rapidity measured in heavy ion collisions at $\sqrt s=2.76$ TeV
at mid rapidity, and momentum integrated \cite{Abelev:2013vea,ABELEV:2013zaa,Abelev:2013xaa,Abelev:2014uua}. The results are
summarized in Table 1.

The baryon number, strangeness and strangeness-electric charge
correlations are obtained from Eq. (11) and from Table 1, as
\begin{align}
 {{\chi_B}\over T^2} &= {1\over{VT^3}} (203.7\pm 11.44)\\
{{\chi_S}\over T^2}  &\simeq {1\over{VT^3}}(504.35\pm 24.14)\\
 {{\chi_{QS}}\over T^2}  &\simeq {1\over{VT^3}}( 178.5\pm 17.14).
 \label{nal}
\end{align}
Particularly interesting are the susceptibility ratios,
\begin{align}\label{nratio}
 {{\chi_B}\over {\chi_S}}&\simeq  0.404\pm 0.028,~~
 {{\chi_B}\over {\chi_{QS}}}\simeq  1.141\pm 0.1266
\end{align}
which  are independent of temperature and  volume.

\begin{table}[!t]
	\begin{tabular}{lcc}
	\hline
	$\langle \pi^{\pm} \rangle$        & & $668.90 \pm 47.50$ \\
	$\langle K^{+} \rangle$            & & $99.67 \pm 8.25$ \\
  $\langle K^{0}_{S}  \rangle$       & & $97.43 \pm 8.00$ \\
  $\langle K^{*}  \rangle$           & & $19.01 \pm 3.18$ \\
  $\langle p \rangle$                & & $30.52 \pm 2.50$ \\
  $\langle \phi \rangle$             & & $12.73 \pm 1.54$ \\
  $\langle \Lambda+\Sigma^0 \rangle$ & & $23.37 \pm 2.50$ \\
  $\langle \Xi^{-} \rangle$          & & $3.34 \pm 0.24$ \\
  $\langle \Omega^{-} \rangle$       & & $0.60 \pm 0.10$ \\

	\hline
	\end{tabular}
 \caption{ALICE data on rapidity distributions at $y=0$ for different
   particle yields in 0-10\% most central Pb--Pb collisions at $\sqrt
   s=2.76$ TeV
   \cite{Abelev:2013vea,ABELEV:2013zaa,Abelev:2013xaa,Abelev:2014uua}.}
\end{table}

 { In Eqs. (12) to (15) the uncertainties of rapidity densities for particles and their
   anti-particles (apart from absorption corrections) were assumed to be fully correlated and therefore
   were added linearly. All remaining errors were treated as being
   independent, thus were added in quadrature. In the calculation of
   the errors of different susceptibility ratios, the partial
   cancelation of errors due to particles which appear both in the
   nominator and denominator has been explicitly taken into account.}

\subsection{Relating LHC data to LQCD }
The net baryon number and strangeness susceptibilities, as well as the
electric charge-strangeness correlations, have been recently
calculated in LQCD at $\mu_B = 0$ for different temperatures
\cite{allton05:_therm_of_two_flavor_qcd,bazavov12:_fluct_and_correl_of_net,borsanyi12:_fluct_of_conser_charg_at,cheng09:_baryon_number_stran_and_elect,Bazavov:2014xya}. The
results are extrapolated to the continuum limit, thus can be directly
compared to heavy ion data.

One expects that a fireball created in heavy ion collisions is of
thermal origin and its properties are governed by statistical QCD, as
quantified by LQCD. If there is a phase change from QGP to the
hadronic phase, then the particle yields and fluctuations of conserved
charges should be established at the chiral, pseudocritical
temperature $T_c$. The value of $T_c$ is well established by LQCD and
coincides within a different discretization scheme of fermionic
action. The value, $T_c=155(1)(8)$ was recently obtained in LQCD with
domain wall fermions \cite{domein}, which preserves all relevant
symmetries of QCD.

The most transparent way to check if the fluctuations of conserved
charges, extracted from ALICE data, are consistent with LQCD at
$T\simeq T_c$, is to compare the ratios from Eq. (\ref{nratio}) to the
corresponding LQCD results.

In Fig. (\ref{fig1}), we compare $\chi_B/\chi_S$, $\chi_B/\chi_{QS}$
and $\chi_S/\chi_{QS}$ ratios with the continuum limit extrapolated
LQCD values at pseudocritical temperature, $T_c=155 $ MeV
\cite{Bazavov:2014xya,bazavov12:_fluct_and_correl_of_net}.  {
  Figure (\ref{fig1}) shows that, within systematic uncertainties,
  there is very good agreement between Pb--Pb collision data from the
  ALICE experiment at the LHC and the LQCD results at $T\simeq 155$
  MeV.}


However, the value of the temperature, at which experimental results
and theory predictions agree, cannot be uniquely determined by
comparing ratios shown in Fig. (\ref{fig1}). This is illustrated in
Figs. (\ref{fig2}) and (\ref{fig3}), where experimental results for
$\chi_B/\chi_S$ and $\chi_B/\chi_{QS}$ from Eq. (\ref{nratio}) are
compared with LGCD predictions at different temperatures
\cite{Bazavov:2014xya}.

The LQCD susceptibility ratios exhibit a rather weak temperature
dependence,  and for $T>0.15$ GeV, are consistent, within statistical
and systematic
uncertainties, with results obtained by using  ALICE data. From Figs. (\ref{fig2}) and
(\ref{fig3}), one can exclude temperatures $T\leq 0.15$ GeV as a
possible range where the saturation of fluctuations in heavy ion data
appears. The upper limit, on the other hand, can be as large as 0.21
GeV.

However, based on different combinations of charge fluctuations and
correlations, it was shown, that at $T>163$ MeV, the LQCD
thermodynamics can not be anymore described by hadronic degrees of
freedom \cite{Karsch:2013naa}. This argument reduces a conceivable window for the
saturation of the net baryon number and strangeness fluctuations to
$0.15<T<0.163$ GeV.


\begin{figure}[!t]
  \vskip -1.0cm\includegraphics[width=3.60in,angle=-90]{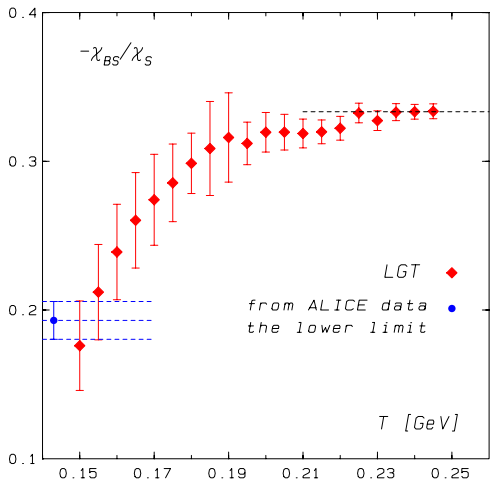}\hskip 0.9cm
 \vskip -1.4cm \caption{The LQCD results  on temperature dependent ratio of baryon-strangeness correlation  $\chi_{BS}$ and
   strangeness susceptibility  from Ref.
   \cite{bazavov12:_fluct_and_correl_of_net}. Also shown is a band
   for the lower limit on this ratio extracted from ALICE data from
    Eqs. (\ref{zall}) and (\ref{upper}). The horizontal line at high-$T$ is an ideal gas value in  a QGP.  }\label{fig4}
 \end{figure}

{ Further constraints on the lower temperature limit for chemical
  freezeout in heavy ion collisions at the LHC can be also obtained by
  considering correlations between strangeness and baryon number,
  $\chi_{BS}$.  Particulary interesting is the ratio
  $\chi_{BS}/\chi_S$, which was proposed as a diagnostic observable
  for deconfinement \cite{koch1}.

The $\chi_{BS}$ correlations are obtained from Eq. (\ref{cors}). Their
upper limit can be expressed by yields of measured strange baryons by
ALICE Collaboration, as
\begin{align}\label{upper}
 -{{\chi_{BS}}\over T^2} & > {1\over{VT^3}} [
   2\langle\Lambda+\Sigma^0\rangle+4\langle\Sigma^+\rangle \\\nonumber
   & +8\langle\Xi\rangle+ 6\langle\Omega^-\rangle ]=97.4\pm 5.8.
\end{align}
Eq. (\ref{upper}) sets only an upper limit for $\chi_{BS}$ since,
e.g., the contributions of strange baryonic resonances decaying into
non-strange baryon and strange meson, like decay of $\Sigma^*\to N\bar
K$, are not included as they are not known experimentally.

In Fig. \ref{fig4} we show the $(-\chi_{BS}/\chi_S)$ ratio obtained in
LQCD by the HotQCD collaboration
\cite{bazavov12:_fluct_and_correl_of_net}.  The LQCD results are
compared with the lower limit, $(-\chi_{BS}/\chi_S)>0.193\pm 0.0127$,
obtained from Eqs.  (\ref{zall}) and (\ref{upper}), and ALICE data
summarized Table 1.  A strong increase of this ratio with temperature,
makes it an ideal observable to fix the temperature in HIC through a
direct comparison of data to LQCD results. From Fig. \ref{fig4}, it is
clear that data are pointing towards temperatures $T>0.15$ GeV. This
supports the conclusion already drawn from Figs. \ref{fig2} and
\ref{fig3}.

The agreement  of the fluctuation ratios extracted from ALICE data
and LQCD in the chiral crossover, seen in
Figs. (\ref{fig1}-\ref{fig4}), supports our assumption, that at the
QCD phase boundary, the second order cumulants and charge correlations
are well approximated by an uncorrelated particle production.}

\vskip 0.2cm
The susceptibilities in Eqs.  (12), (13) and
(14)  { can also be used to obtain information on  the volume of the fireball for one unit of rapidity},

 \begin{align}\label{v1}
V_{\chi_B}= {{ 203.7\pm 11.44}\over {T^3(\chi_B/T^2)_{LQCD} }}, ~~V_{\chi_S}= {{504.35\pm 24.14 }\over {T^3(\chi_S/T^2)_{LQCD} }}
\end{align}
and
\begin{align}\label{v2}
V_{\chi_{QS}}= {{ 178.5\pm 17.14 }\over {T^3(\chi_{QS}/T^2)_{LQCD} }}
\end{align}
Clearly, if $\chi_B$, $\chi_S$ and $\chi_{QS}$ are established in a
common fireball, then not only the temperature, but also the
corresponding volumes, $V_{\chi_B}$, $V_{\chi_S}$ and $V_{\chi_{QS}}$
must be equal.

In Fig.(\ref{figv}), we show the temperature dependence of volume
parameters obtained from Eqs. (\ref{v1}) and (\ref{v2}). There is a
clear decrease of volume with temperature, which is needed to
reproduce LQCD susceptibilities. For a given temperature, the volume
of the fireball is extracted as overlap of all $V_{\chi_B}$,
$V_{\chi_S}$ and $V_{\chi_{QS}}$.

The volume parameters from Fig. (\ref{figv}), together with the total
number of particles in the final state $\langle N_{t}\rangle$, is used
to calculate the density of particles in a collision fireball,
$n(T)=\langle N_{t}\rangle/V(T)$.

We calculate the total number of particles per unit of rapidity at
mid-rapidity in central Pb-Pb collisions
at $\sqrt s =2.76$ TeV, as follows

\begin{align}\label{neutral}
\langle N_t\rangle= & 3\langle\pi\rangle+ 4\langle K\rangle
+4\langle p \rangle +2\langle\Lambda+\Sigma^0\rangle
+4\langle\Sigma\rangle \\\nonumber &+4\langle
\bar \Xi\rangle +2\langle\bar \Omega\rangle,
\end{align}
which gives $\langle N_t\rangle=2486\pm 146$

In Fig. (\ref{figv}), we show the corresponding density of particles,
$n(T)$. Clearly, due to deconfinement, there is a limiting temperature
and corresponding density, above which the fireball constituents can
not be hadronic anymore.

 In percolation theory of objects of (eigen-)volume $V_0$, there is a
 critical density, $n_c^{per}=1.22/V_0$ beyond which the objects start
 to overlap \cite{castorina}. Relating percolation to deconfinement
 \cite{castorina}, one can estimate the critical particle density in
 the hadronic phase.  { Considering hadrons as objects of volume
   $V_0=(4/3)\pi R_0^3$, with $R_0=\sqrt {\langle r_p^2\rangle}$ and
   $\langle r_p^2\rangle=0.67\pm 0.02$ being the mean squared strong
   interaction radius of the proton \cite{proton}\footnote{{ This
       hadronic radius of the proton is somewhat smaller than its recently
       obtained charge radius, $r_p^E= 0.84\pm 0.01$
       \cite{meissner}.}}, one gets, $n_c^{per}\simeq 0.53\pm 0.024$
   fm$^{-3}$.  The central $n_c^{per}$ value is also marked in
   Fig. (4). 

\begin{figure}[!t]\label{nfig4}
\vskip -1.1cm \hskip -0.3cm\includegraphics[width=3.60in,angle=-90]{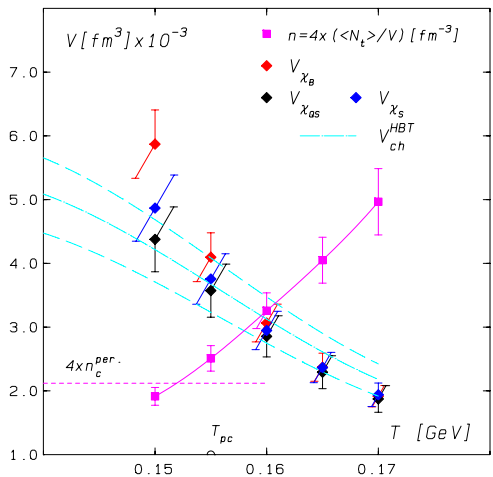}\hskip 0.9cm
\vskip -1.2cm  \caption{Volume calculated from Eqs. (\ref{v1}) and
   (\ref{v2}).  Also shown are results for particle density at
   corresponding temperature and the
   critical density in percolation theory, as well as, { the chemical freezeout volume
$V_{ch}^{HBT}$,   extracted from the HBT data at thermal freezeout  and rescaled  to higher $T$ within   3D-hydrodynamics. }}\label{figv}
 \end{figure}

Remarkably, this critical percolation density appears at $T= 152\pm 1$
MeV, thus within systematic uncertainties, is consistent with the
transition temperature obtained from LQCD.  Clearly, the value of $T$
at $n_c^{per}$ strongly depends on $R_0$. The lower limit of
$R_0\simeq 0.67$ fm corresponds to $T\simeq 163$ MeV, since above this
temperature, the LQCD thermodynamics is not anymore described by the
hadronic degrees of freedom. This lower value of $R_0$ coincides,
within error, with a measured charge radius of the pion, $\sqrt {\langle
r_\pi^2\rangle} =0.657\pm 0.012$ \cite{pion}.  Consequently, the
percolation of pions and protons appears at temperatures which overlap
with the QCD chiral crossover.}

Some limitations on the volume of the fireball, thus also on
temperature, can  be imposed from the HBT interferometry measurements.
The Hanbury-Brown-Twiss (HBT) analysis of multiparticle production
processes is becoming a widely used technique in heavy ion
collisions. It provides information on the space-time evolution of an
excited strongly interacting system produced in high energy
collisions.

The HBT volume, can be obtained from the product of the longitudinal
$R_l$, outward $R_o$ and the sideward $R_s$ radius, as
$V_{HBT}=(2\pi)^{3/2}R_lR_oR_s$, if the $R_i$ are rms values of
Gaussian distributions. From the first measurement of
two-pion Bose-Einstein correlations in central Pb-Pb collisions at
$\sqrt{s_{NN}} = 2.76$ TeV at the LHC by ALICE Collaboration \cite{AHBT},
one gets  $V_{HBT}=4800\pm 580  fm^3$ for centrality (0-5\%).
{We note, however, that $V_{HBT}$ is, {{in general considered as}} the volume at
thermal freeze-out.  {Thus, the fireball volume at chemical freezeout $V_{ch}^{HBT}(T)$ is smaller than the $V_{HBT}$ introduced above due to the expansion of the system between chemical and thermal freeze-out.}
 {To connect}  $V_{CH}^{HBT}(T)$ with $V_{HBT}$ involves model
assumptions which we discuss briefly below.

Furthermore, $V_{HBT}$ is not representing the source size, but only
  { the
  volume of the homogeneity region at the last interaction}. Following
the procedure developed in \cite{Adamova:2002ff} we estimate that the
true  { thermal} freeze-out volume per unit of rapidity exceeds
the $V_{HBT}$ value above by a factor of 1.28 for thermal freeze-out
at $T=155$ MeV, by 1.47 at $T = 120$ MeV and by 1.63 at $T = 100$ MeV.  We
further note that $V_{HBT}$ grows with the charge particle
multiplicity \cite{AHBT}, as expected from the fireball volume at
chemical freeze out. As a consequence we use the volume appropriate
for (0 - 10\%) centrality and correct with the above factors.

The relevant corrected volumes  are then
$V_{ch} = 5510 \pm 670$ fm$^3$ for T = 155 MeV,
$V_{ch} = 6340 \pm 770$ fm$^3$ for T = 120 MeV and
$V_{ch} = 7050 \pm 850$ fm$^3$ at T = 100 MeV.
For thermal freeze-out at 155 MeV, i.e. close to the chiral crossover
temperature, there is no further extrapolation needed, and the minimal
corrected volume is $V_{ch}^{HBT} = 4840 $ fm$^3$.
{As seen in Fig. \ref{figv}, this chemical freezeout volume
  is within uncertainies comparable with that  extracted from the LQCD analysis at
  temperature $T = 155$ MeV.}

For thermal freeze-out at $T=100$ and $T =120$ MeV, one can  calculate  the
volume decrease with increasing   temperature, i.e towards the chiral
crossover temperature,
 {by employing models for expansion dynamics in heavy ion collisions.}
{ We have here adopted the  3D-hydrodynamics approach with
  initial conditions appropriate for LHC energy and calculated in the MC Glauber
model to extract the relative change of  the fireball volume with temperature \cite{hydro}.}

{In Fig. 5 we show the $T$-dependent fireball volume
  $V_{ch}^{HBT}(T)$ obtained by rescaling the above kinetic freezeout
  volume at $T=100$ MeV with the factor obtained from the
  3D-hydrodynamics for a temperature range between 140 and 170 MeV.
  Starting from the kinetic freezeout volume at $T=120$ MeV leads to
  very similar results.

As seen in Fig. (\ref{figv}), the $V_{ch}^{HBT}(T)$ coincides with the
volume extracted from LQCD and ALICE data, in the chiral crossover
region. This by itself is a non-trivial observation.
At the same time, such comparison does not restrict the value
of the chemical freezeout temperature at LHC energy.}}

{ From the comparison of different fluctuation ratios extracted
  from ALICE data and LQCD results, one concludes that, in heavy ion
  collisions at the LHC, the 2nd order fluctuations are of thermal
  origin and saturate at LQCD values at temperature $T>0.15$ GeV.
  This together with the LQCD observation, that at $T>0.163$ GeV, the
  fluctuations of conserved charges can not be anymore described by
  the hadronic degrees of freedom implies that $0.15<T_c<0.163$ is the
  most likely temperature range for particle freezout at the LHC.  In
  this temperature window, which overlaps with the chiral crossover
  temperature, the $T$-correlated fireball volume per unit rapidity is
  obtained from Fig. 5 as $5000 > V \geq 3000$ fm$^3$. This range of
  volumes is also consistent with that extracted from the HBT
  measurement and extrapolated to higher temperatures within
  3D-hydrodynamics \cite{hydro}.}

A recent analysis of particle yields in heavy ion collisions at the
LHC, within the thermal model, has shown that $T\simeq 156$ MeV and
$V\simeq 5300$ fm$^3$, reproduce all yield data
\cite{Stachel:2013zma}. This temperature value agrees well with the present
analysis. The somewhat larger volume in Ref. \cite{Stachel:2013zma} appears, since repulsive
interactions of particles were included in the analysis. In this case,
particle densities are reduced, and to reproduce measured yields, a
larger volume is required.

\section{Concluding remarks}\label{sec:summary}
We have proposed a method to construct the net baryon number and
strangeness susceptibilities as well as correlations between
electric charge and strangeness from experimental data of the ALICE
Collaboration, taken in Pb-Pb collisions at $\sqrt{s_{NN}}$=2.76 TeV.

 { Using this approach, we have shown that fluctuations and
   correlations derived from ALICE data at the LHC are consistent with
   LQCD predictions in the temperature window, $0.15 < T_c\leq 0.163$
   GeV, which overlap with the chiral crossover.  In this temperature
   interval, the fireball volume per unit rapidity corresponds to
   $5000 > V \geq 3000$ fm$^3$.

 Such a direct agreement between experiment and LQCD lends strong
 support to the notion that the fireball created in central
 nucleus-nucleus collisions at the LHC is of thermal origin and
 exhibits characteristic properties expected in QCD at the transition
 from a quark-gluon plasma to a hadronic phase.}

We have discussed uncertainties in the determination of temperature and
volume of the fireball at the LHC.  We have also discussed possible constraints on the
parameters originating from pion interferometry measurements and
percolation theory.

The analysis presented here provides the first direct link between
LHC heavy ion data and predictions from LQCD. This was possible since, at the LHC, the
conditions of charge neutrality in the fireball directly
match that in LQCD calculations. In addition, the constructed
susceptibilities and correlations contain contributions from all
charged hadrons integrated over the full momentum range. This is essential
and necessary to make a successful comparison of data to the first
principle LQCD calculations.

{ Finally, our method is based on the assumption that the probability
distribution of baryons at LHC energy is close to a Skellam
distribution. The probability distribution for net baryon number
production at LHC energy can be directly obtained from measurements of
protons and anti-protons. Since we are interested in results at
midrapidity, issues related to isospin can be safely neglected and
proton and neutron  mean numbers should be equal. Furthermore, corrections
due to baryon number conservation should be negligible since near
midrapidity the baryon rapidity distribution is very close to
flat. Since the present method only relies on the second moment of the
distribution, very high statistics is not needed and a  typical  $10^6$ central
collisions should be sufficient, implying that our assumption can be tested
experimentally in the near future.}

\acknowledgments
We acknowledge  stimulating discussions with Bengt Friman and Frithjof Karsch.
We also acknowledge fruitful critical comments from Volker Koch.
  We are  grateful to  Vladimir Shapoval and Yuri Sinyukov for fruitful discussions and for providing results from hydrodynamics calculations.  This work was partly supported by the Polish Science Foundation (NCN), under
 Maestro grant DEC-2013/10/A/ST2/00106.

\end{document}